\documentclass[final,5p,times,twocolumn]{elsarticle}
\usepackage{amssymb}
\usepackage{tabularx}
\usepackage{amsmath}
\usepackage{amsmath,bm} % 导言区添加 bm 包
\usepackage{booktabs} % 专业表格宏包
\usepackage{graphicx} % 必须加载的宏包
\usepackage{multirow} % 用于多行合并单元格
\usepackage{array}    % 用于列格式调整
\usepackage{caption}
\usepackage{makecell}    % 多行表头
\usepackage{siunitx}     % 数字对齐
\usepackage[colorlinks]{hyperref} % 使引用可点击
\usepackage{cleveref}             % 智能引用
\usepackage[normalem]{ulem} % 防止与其他包冲突
\journal{Nuclear Physics B}

\begin{document}

\begin{frontmatter}


\title{TPPR: APT Tactic / Technique Pattern Guided Attack Path Reasoning for Attack Investigation\tnoteref{}}
\author{Name\corref{cor1}\fnref{label2}}
\ead{email address}
\ead[url]{home page}
\fntext[label2]{}
\cortext[cor1]{}
\affiliation{organization={},
            addressline={},
            city={},
            postcode={},
            state={},
            country={}}
\fntext[label3]{}
% \title{} %% Article title


\author{Qi Sheng,Zhejiang University of Technology} %% Author name

\affiliation{organization={},%Department and Organization
            addressline={}, 
            city={},
            postcode={}, 
            state={},
            country={}}

%% Abstract
\begin{abstract}
Provenance analysis based on system audit data has emerged as a fundamental approach for investigating Advanced Persistent Threat (APT) attacks. Due to the high concealment and long-term persistence of APT attacks, they are only represented as a minimal part of the critical path in the provenance graph. While existing techniques employ behavioral pattern matching and data flow feature matching to uncover latent associations in attack sequences through provenance graph path reasoning, their inability to establish effective attack context associations often leads to the conflation of benign system operations with real attack entities, that fail to accurately characterize real APT behaviors. We observe that while the causality of entities in the provenance graph exhibit substantial complexity, attackers often follow specific attack patterns—specifically, clear combinations of tactics and techniques to achieve their goals. Based on these insights, we propose TPPR, a novel framework that first extracts anomaly subgraphs through abnormal node detection, TTP-annotation and graph pruning, then performs attack path reasoning using mined TTP sequential pattern, and finally reconstructs  attack scenarios through confidence-based path scoring and merging. Extensive evaluation on real enterprise logs (100+ million events) and DARPA TC dataset demonstrates TPPR's capability to achieve 99.9\% graph simplification (700,000 to 20 edges) while preserving 91\% of critical attack nodes, outperforming state-of-the-art solutions (SPARSE, DepImpact) by 63.1\% and 67.9\% in reconstruction precision while maintaining attack scenario integrity.

\end{abstract}


%%\item Research highlight 1
%%\item Research highlight 2

%% Keywords

\end{frontmatter}



\section{Introduction}
Advanced Persistent Threats (APTs) represent highly stealthy and prolonged cyber attacks, meticulously orchestrated to exfiltrate sensitive data or compromise critical infrastructure through a multi-stage process \cite{C1}. As attack vectors grow increasingly sophisticated, traditional security defenses often prove inadequate, posing significant risks to enterprises and institutions.

To address this challenge, existing approaches introduce provenance technique to enhance APT detection and investigation. By collecting and converting audit logs into a graph structure (called provenance graph), centered on system entities (e.g., processes, files) and their interaction events (e.g., create processes, read files), the provenance technique constructs a comprehensive and bird’s-eye view of system activities, facilitating deeper causal analysis of APTs. Based on this technique, APT detection methods have demonstrated effective in precisely identifying threats \cite{C2,C3,C4,C5}. However, these methods are constrained to pinpointing isolated malicious events, while struggling to reconstruct the complete APT attack paths that typically involve multiple interconnected attack techniques and tactics. This limitation leads to a large amount of false alarms on one hand, and impedes security analysts from fully grasping the scope and implications of an attack on the other hand. Consequently, the development of expeditious mitigation strategies is hampered.

Attack path reasoning, a technique aims to reconstruct attack scenarios (composed of multiple attack paths) and reduce false alarms by uncovering causal correlations over malicious system events, has become a critical auxiliary component for APT detection \cite{C21}. Due to the gigantic size of provenance graphs, manually tracing such correlations between system events is prohibitively time-consuming and labor-intensive. In response, recent studies have moved towards automated APT attack path reasoning, which generally falls into three strategies. First, the anomaly based approach detects events that deviate from normal behavior patterns and then reduces the size of provenance graphs by retaining only abnormal nodes and edges \cite{C6}, so that the attack path reasoning can be efficiently performed. However, since the anomaly based approach can only perform binary malicious/benign classification on events, it cannot distinguish which malicious events belong to the same attack campaign. Second, the threat intelligence based approach uses Cyber Threat Intelligence (CTI) to frame the attack path reasoning task as an inexact graph pattern matching problem. Although it demonstrates strong reasoning performance, its adaptability to unknown attacks remains constrained, as CTIs only encompass previously observed attacks \cite{C7}. Third, the label propagation based approach employs manually crafted rules for attack path reasoning, offering better flexibility and interpretability. Nevertheless, its reliance on expert experience proves inadequate in accommodating the diversity and variability of emerging attacks \cite{C5,C9}, thereby hindering its generalization.

%% Labels are used to cross-reference an item using \ref command.

%% Use \subsection commands to start a subsection.
\paragraph{\textbf{Key Insight}} 
Through a comprehensive analysis of diverse APT scenarios, we observe that, despite the continual evolution of adversarial methodologies, most APT attack paths exhibit consistent and structured patterns of techniques and tactics (e.g.,  as defined by ATT\&CK framework) \cite{C10}. Specifically, APT attack campaigns with common goals typically progress through similar ATT\&CK tactical stages. Along the way, adversaries might also employ auxiliary tactics to ensure the successful execution of each primary step. For example, consider two ransomware campaigns conducted by Conti and AvosLocke, both of which utilize ATT\&CK tactics including Initial Access, Defense Evasion, Lateral Movement, and Impact \cite{C11,C12}. Such recurring patterns across varied attack campaigns underscore the structured and systematic nature of APT operations, which can, in reverse, serve as a foundation for attack path reasoning and proactive threat analysis.

To this end, we propose TPPR, an APT \uline{\textbf{T}}actic/Technique \uline{\textbf{P}}attern Guided Attack \uline{\textbf{P}}ath \uline{\textbf{R}}easoning framework, which correlates isolated malicious events by mining and utilizing ATT\&CK attack tactic/technique patterns. First, since the direct reasoning over large-scale provenance graphs is computationally prohibitive, we reduce graph complexity by employing node-level anomaly detection and edge compression strategies, condensing the provenance graph into a focused anomaly subgraph that preserves only suspicious semantics. Second, abnormal nodes within these compacted subgraphs are annotated with their exploited tactics and techniques, enhancing semantic clarity for downstream analysis. Third, building upon the ATT\&CK framework, we extract sequential patterns of attack tactics and techniques as the knowledge base, and design an attack path inference algorithm that integrates graph walking and threat scoring mechanisms to prune anomaly subgraph into a concise attack summary graph. In summary, the main contributions of this paper are as follows:

1.We propose TPPR, a novel framework that utilizes attack tactic/technique patterns derived from ATT\&CK for APT attack path reasoning on provenance graphs.

2.We introduce a hybrid method combining abnormal node detection and attack tactic/technique recognition to construct meaningful anomaly subgraphs.

3.We develop an APT attack path inference algorithm guided by attack tactic/technique patterns, and integrate heuristic graph walking and threat scoring mechanisms to generate attack summary graphs.

4.Experimental evaluations on publicly available datasets and our own simulated attack cases demonstrate the effectiveness of TPPR in APT attack path reasoning.
%% Use \subsubsection, \paragraph, \subparagraph commands to 
%% start 3rd, 4th and 5th level sections.
%% Refer following link for more details.
%% https://en.wikibooks.org/wiki/LaTeX/Document_Structure#Sectioning_commands
\section{Background and Motivation}
\subsection{Provenance Graph}
Analyzing system audit data to uncover attack paths is inherently challenging due to the overwhelming volume and scattered nature of system events. To solve this problem, current research leverages provenance techniques to structure discrete events into a provenance graph, which causally links each attacking stage from the initial intrusion to the final compromise, thereby enabling a more systematic analysis. A provenance graph is a directed and acyclic structure, defined as $G$ = ($E$, $V$). Here, $V$ represents the set of attributed system entities (i.e., files, processes, sockets), while $E$ denotes the set of interactions between system entities (e.g., read files). Each edge $e$ = ($u$, $v$, $a$, $t$) captures a system event, where $u$ is the subject, $v$ is the object, $a$ is the type of interaction, and $t$ is the timestamp.

By constructing a provenance graph, security analysts can efficiently identify the key intrusion points through backward tracing and discover potential ramifications via forward tracing, significantly streamlining the reasoning and analysis of multi-stage attacks. The considered system entities and system events are listed in \Cref{tab:tb1}.

% Use a table environment to create tables.
% Refer following link for more details.
% https://en.wikibooks.org/wiki/LaTeX/Tables
\begin{table}[htbp]
  \centering
  \caption{The monitored system events.}
  \label{tab:tb1}
  \small  % 使用小号字体
  \begin{tabular}{@{}>{\raggedright\arraybackslash}p{2.2cm}
                   >{\raggedright\arraybackslash}p{2.6cm}
                   >{\raggedright\arraybackslash}p{2.8cm}@{}}
    \toprule
    \textbf{Events} & \textbf{Operations} & \textbf{Attributes} \\
    \midrule
    Process→File & read,write,chmod & PID,Name,Path,Cmd \\
    Process→Process & start,end,execve,clone & PID,Name,Cmd \\
    Process→Socket & sendto,recvfrom,copy & PID,Name,Cmd,IP \\
    \bottomrule
  \end{tabular}
\end{table}

\subsection{Motivating Example}%2.2
\begin{figure*}[t]
  \centering
  \includegraphics[width=\textwidth, height=8cm, keepaspectratio]{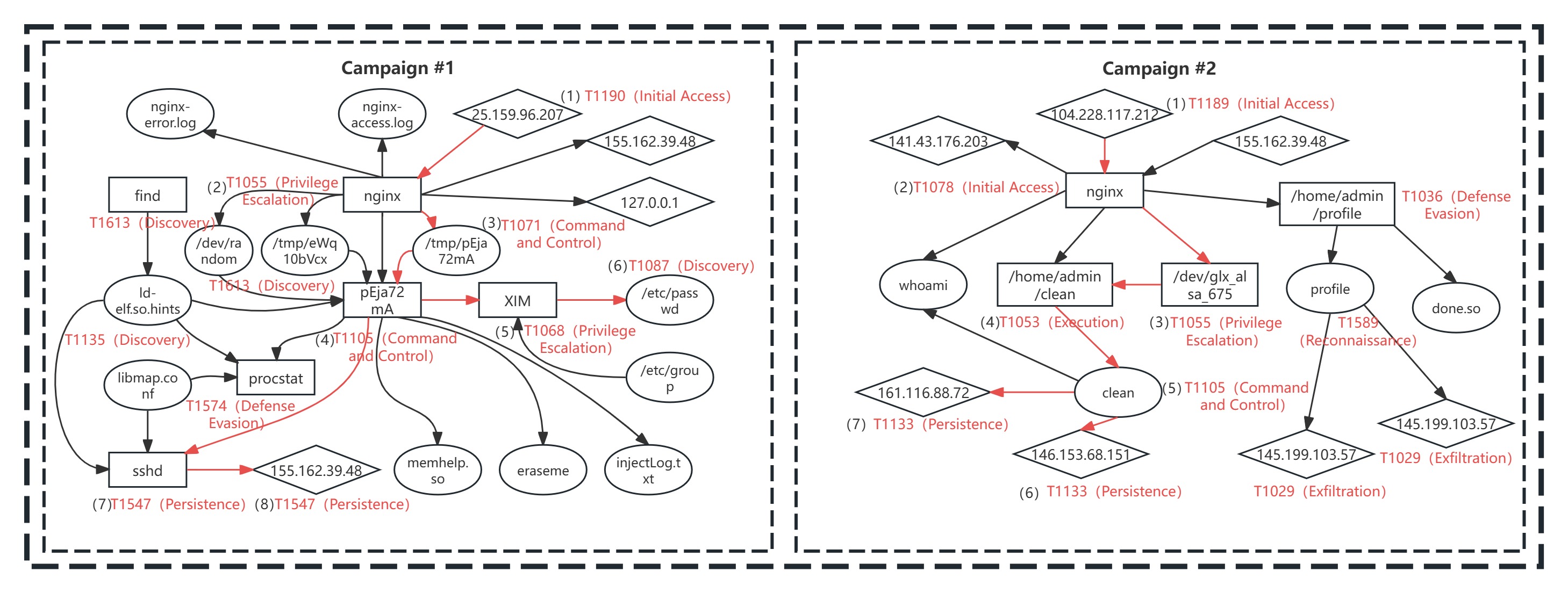} % 替换为你的文件名
  \caption{A motivating example.}
  \label{fig:example1}
\end{figure*}

\label{subsec:2.2}
We use the APT campaigns from DARPA TC Engagement 3 as a motivating example. As illustrated in \Cref{fig:example1}, two similar APT attack campaigns are presented. In Campaign \#1, the attacker exploits Nginx with a malformed HTTP request to launch an attack against CADETS. This exploit results in a drakon implant running in Nginx memory, and a shell then connects via HTTP to the operator console. The attacker downloads drakon implant executable to the target disk and executes it with elevated privileges, resulting in a new drakon implant process with root privileges connecting out to the adversary’s console. In Campaign \#2, the attacker uses a malicious ad server via a website. The exploit results in the drakon implant running in memory in the Firefox process with a connection out to the attacker operator console. Then, they write the implant to disk, execute it with root privileges through escalation to gain persistent high-level access, regain access when the victim revisits the site, and deploy additional files while maintaining an open C2 channel for ongoing operations.

To facilitate analytical visualization, we abstract the attack campaigns into anomaly subgraphs (as shown in \Cref{fig:example1}), where nodes represent detected anomaly instances annotated with their corresponding ATT\&CK attack tactic/technique. Red arrows denote the attack paths. Through empirical observation of the attack campaigns, we identify consistent tactical/technical sequential patterns governing attack paths. For instance, Campaign \#1 follows the attack paths “T1190 (Initial Access) → T1055 (Privilege Escalation) → T1071 (Command \& Control) → T1105 (Command \& Control) → T1068 (Privilege Escalation) → T1087 (Discovery)”, and Campaign \#2 follows the attack paths “T1189 (Initial Access) → T1078 (Initial Access) → T1055 (Privilege Escalation) → T1053 (Execution) → T1105 (Command \& Control) → T1133 (Persistence)”.

It can be found that the two attack campaigns share similar subsequences of attack tactics, which reveal the attackers’ adherence to logically structured tactical sequences to achieve objectives. In addition, entities outside the attack paths demonstrate incoherent attack tactic subsequences, lacking logical progression. This insight motivates our proposition, i.e., mining attack tactic/technique sequential patterns to filter out noises in provenance graphs and probabilistically guide attack scenario reconstruction.

\subsection{Threat Model}
We assume the integrity of both the system kernel and the auditing, and presume audit logs are free from tampering or manipulation, thereby ensuring that audit logs accurately and reliably reflect system activities as recorded by the provenance tracker. Attacks that compromise kernel-level integrity, exploit side-channel vulnerabilities, or deliberately target the auditing infrastructure fall outside the scope of this work. We also exclude mimicry attacks, where an adversary deliberately evades intrusion detection systems by crafting a sequence of seemingly benign events. Consistent with \cite{C16}, our approach instead focuses on attack scenario characterization through identification of relevant events and attack path inference.

\begin{figure*}[t]
  \centering
  \includegraphics[width=\textwidth, height=8cm, keepaspectratio]{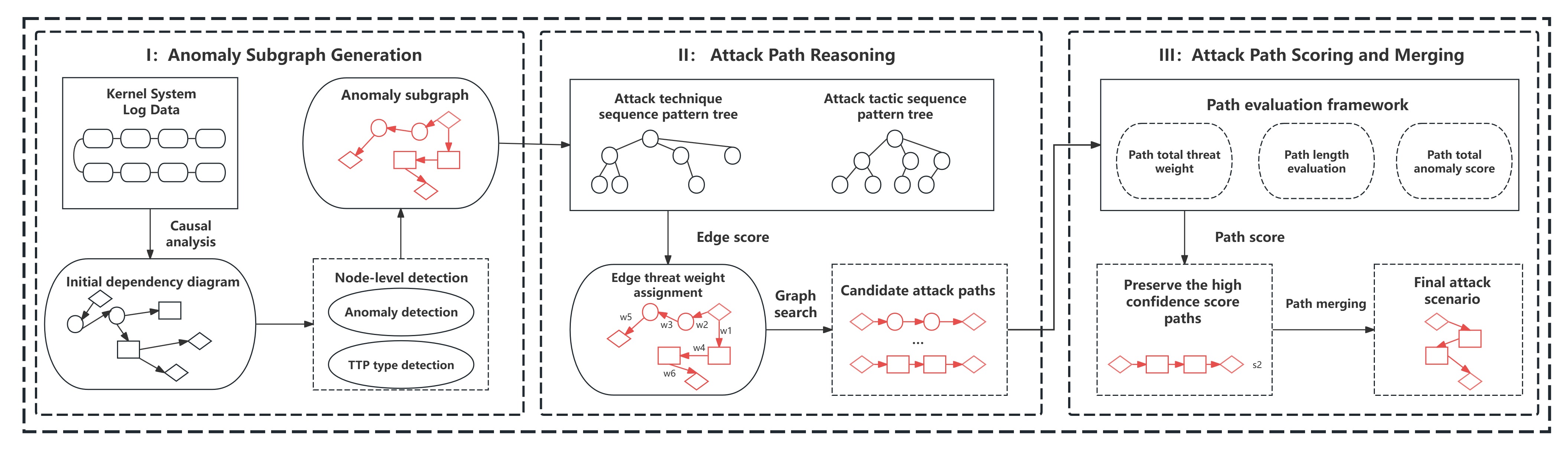} % 替换为你的文件名
  \caption{The architecture of TPPR.}
  \label{fig:example2}
\end{figure*}
%% Use figure environment to create figures
%% Refer following link for more details.
%% https://en.wikibooks.org/wiki/LaTeX/Floats,_Figures_and_Captions
% \begin{figure}[t]%% placement specifier
% %% Use \includegraphics command to insert graphic files. Place graphics files in 
% %% working directory.
% \centering%% For centre alignment of image.
% \includegraphics{example-image-a}
% %% Use \caption command for figure caption and label.
% \caption{Figure Caption}\label{fig1}
% %% https://en.wikibooks.org/wiki/LaTeX/Importing_Graphics#Importing_external_graphics
% \end{figure}
\section{System Design}

\subsection{Overview}%3.1
\Cref{fig:example2} presents the architecture of TPPR, which consists of three phases, i.e., anomaly subgraph mining, attack path reasoning, and attack path scoring and merging. Each phase is outlined below.

1.Anomaly Subgraph Mining: First, given a provenance graph, abnormal nodes are detected using an anomaly detection model. Second, by performing TTP(Tactic, Technique, and Procedure) recognition, the identified abnormal nodes are annotated with their corresponding ATT\&CK attack tactics and techniques. Third, the original provenance graph is pruned according to the abnormal nodes, producing a simplified anomaly subgraph.

2.Attack Path Reasoning: First, it extracts the sequential patterns of attack tactics and techniques through the analysis of CTI reports with TTP annotations. Then, a graph walk algorithm is applied to identify potential candidate attack paths based on the extracted tactic and technique sequential patterns.

3.Attack Path Scoring and Merging: First, threat scores of each candidate attack path are computed by integrating a comprehensive set of risk factors, such as the matching degree with attack tactic/technique patterns, the connectivity (in-out degrees) and potential influence of each node, and the anomaly scores assigned to each node by the detection system. Then, the candidate paths are screened, with those assigned higher threat scores being prioritized. Finally, the candidate attack paths with common nodes are merged to reduce redundancy.

\subsection{Anomaly Subgraph Mining}%3.2
\label{subsec:3.2}
\subsubsection{Abnormal Node Detection}%3.2.1
\label{subsec:3.2.1}
With the increasing complexity of systems and applications, the provenance graph grows excessively intricate and large, making direct threat analysis impractical. To address this, we employ abnormal node detection to pinpoint suspicious nodes and their associated interactions, while filtering out extraneous elements. Then, the size and complexity of provenance graphs can be significantly reduced, enabling more efficient and focused analysis.

Given the presence of malicious nodes concealed among tens of thousands of benign entities within provenance graphs, manually labeling each attack-related node is virtually infeasible \cite{C18}. Consequently, building a binary classification model via supervised learning proves to be impractical. Following existing studies \cite{C19,C20,C21}, we construct an abnormal node detection model through unsupervised learning, i.e., the model is exclusively trained on benign datasets and detects anomalies based on large deviations from the normal patterns.

Specifically, we first initialize the features of each node in the provenance graph as a 20-dimensional feature vector representing the frequency statistics of all its incoming and outgoing edge types (as shown in \Cref{tab:tb1}). Since the initial feature vector cannot capture long-distance causal relationships, we use a graph neural network (GNN) to further aggregate the high-order interaction features, where each GNN layer aggregate adjacent neighbor information. By stacking multiple GNN layers, distant interaction semantics can be embedded. As illustrated in Eq.\eqref{eq:1}, the final layer of the feature matrix E(t) is utilized for the detection of abnormal nodes.

\begin{equation}
    {E}^{(t)}= GNN({W}^{(t-1)},G,{E}^{(t-1)})
    \label{eq:1}  % 设置标签（建议用 eq: 前缀）
\end{equation}

After that, we input the final embedding vectors of all the nodes into an outlier detection algorithm to detect abnormal nodes. iForest has been proven to be an effective outlier detection algorithm. Unlike algorithms that rely on indicators such as distance or density, iForest employs random feature selection, which enhances its sensitivity to critical features. Additionally, it handles high-dimensional data well and avoids the negative impact of high-dimensional distance calculations. These strengths motivate the use of iForest for detecting abnormal nodes within provenance graphs.

\subsubsection{TTP Recognition}%3.2.2
\label{subsec:3.2.2}
By recognizing the TTP type of each abnormal node, we can map associated events to specific tactics/techniques in the ATT\&CK framework, thereby enabling the attack path analysis via tactic/technique patterns. We perform node-level TTP recognition by adopting TREC \cite{C19}, a deep learning-based TTP detection framework.

Specifically, TREC selects highly connected nodes from abnormal nodes as seed nodes, builds a technical subgraph (TSG) through depth-first search, and uses heterogeneous meta-paths to encode them into low-dimensional vectors. Then, a few-shot learning method based on Siamese networks is employed to train a TTP recognition model, achieving TTP classification through distance measurement in feature space. All abnormal nodes are processed in this manner to annotate their corresponding TTP labels.

\begin{figure}[t]
  \centering
  \includegraphics[width=\columnwidth]{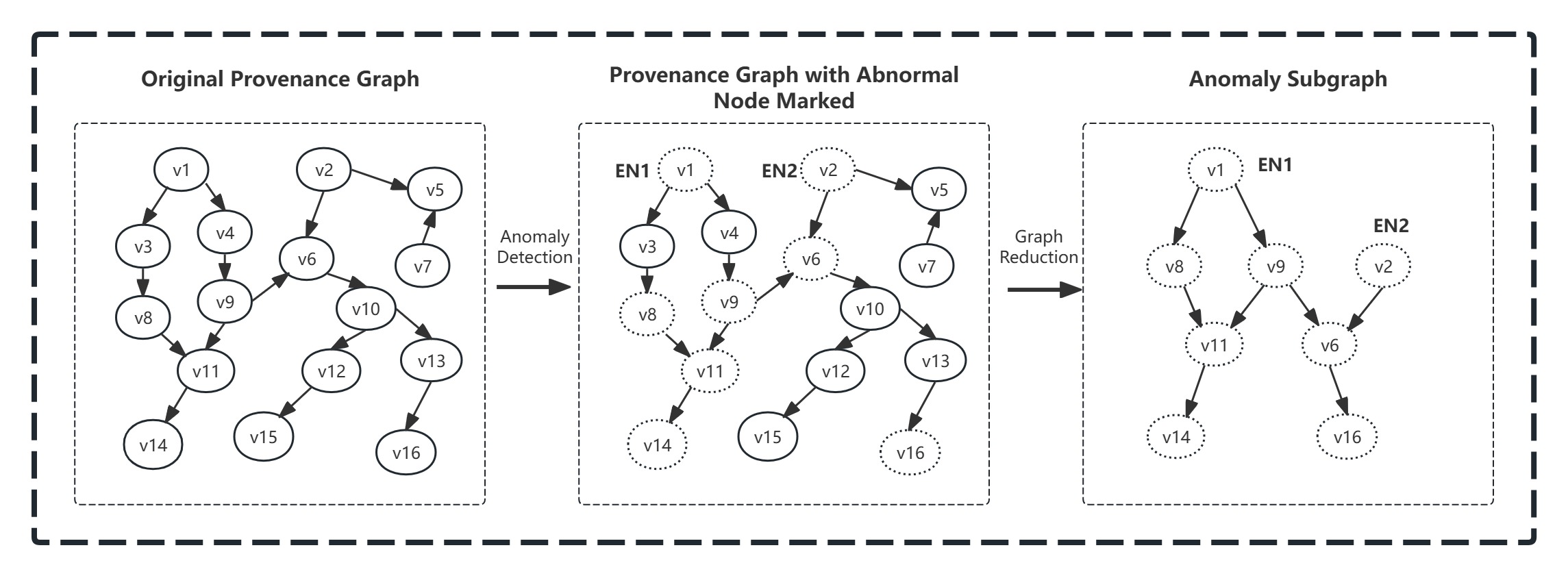} % 替换为你的文件名
  \caption{Reduction of edges between abnormal nodes.}
  \label{fig:example3}
\end{figure}
\subsubsection{Provenance Graph Compression}%3.2.3
Although the investigation scope has been narrowed down to abnormal nodes and their correlated neighboring nodes, analyzing the attacker's intentions is still challenging due to the complex and diverse system events and numerous neighboring correlations.

To further streamline the process, we firstly initialize the entry node set ($ENS$) by selecting all detected abnormal nodes $EN_{k}$ without anomaly predecessors. For each $EN_{k}$ in $ENS$, we perform a depth-first search traversal along dependency edges to enumerate all primitive paths. During the stepwise traversal of paths originating from $EN_{k}$—nodes along the path are iteratively processed until a subsequent abnormal node $AN_{k+1}$ is encountered. At this point, non-anomalous intermediate nodes are pruned while preserving the directed edge $AN_{k}$ →$AN_{k+1}$. This path compression iterates recursively until no subsequent abnormal nodes are reachable in the current path. This iterative path compression culminates in the generation of an anomaly subgraph representing the essential topological structure of attack progression.

We give an example in \Cref{fig:example3}, where the provenance graph undergoes initial anomaly detection with nodes classified as anomalous (dashed) or benign (solid). Abnormal nodes $v_{1}$ and $v_{2}$ without anomalous predecessors are selected as entry points for depth-first traversal, yielding seven original paths: 
$p_{1} = v_{1} \to v_{3} \to v_{8} \to v_{11} \to v_{14}$, $p_{2} = v_{1} \to v_{4} \to v_{9} \to v_{11} \to v_{14}$, 
$p_{3} = v_{1} \to v_{4} \to v_{9} \to v_{6} \to v_{10} \to v_{12} \to v_{15}$, $p_{4} = v_{1} \to v_{4} \to v_{9} \to v_{6} \to v_{10} \to v_{13} \to v_{16}$, 
$p_{5} = v_{2} \to v_{6} \to v_{10} \to v_{12} \to v_{15}$, $p_{6} = v_{2} \to v_{6} \to v_{10} \to v_{13} \to v_{16}$, $p_{7} = v_{2} \to v_{5}$. Through non-anomalous intermediary node pruning, these paths are condensed to: 
$p'_{1} = v_{1} \to v_{8} \to v_{11} \to v_{14}$, $p'_{2} = v_{1} \to v_{9} \to v_{11} \to v_{14}$, $p'_{3} = v_{1} \to v_{9} \to v_{6}$, 
$p'_{4} = v_{1} \to v_{9} \to v_{6} \to v_{16}$, $p'_{5} = v_{2} \to v_{6}$, $p'_{6} = v_{2} \to v_{6} \to v_{16}$, $p'_{7} = v_{2}$, 
generating an anomaly subgraph comprising nodes $\{v_{1}, v_{2}, v_{6}, v_{8}, v_{9}, v_{11}, v_{14}, v_{16}\}$. In this process, large volumes of unrelated nodes are filtered out, and possible anomalous propagation paths are highlighted, significantly reducing graph complexity while preserving the topological integrity of the attack path.

\subsection{Attack Path Reasoning}%3.3
\label{subsec:3.3}
\subsubsection{Attack Tactic/Technique Pattern Mining}%3.3.1
As discussed in \Cref{subsec:2.2}, attack tactic/technique patterns can be utilized to effectively correlate isolated abnormal nodes and form attack paths by filtering out irrelevant nodes that deviate from expected adversarial behaviors. To achieve this, we extract attack tactic/technique patterns via two steps: collecting CTI reports annotated with TTPs, and constructing the attack tactic/technique sequential pattern tree (abbreviated as ATT-SPT).

\begin{figure}[t]
  \centering
  \includegraphics[width=\columnwidth]{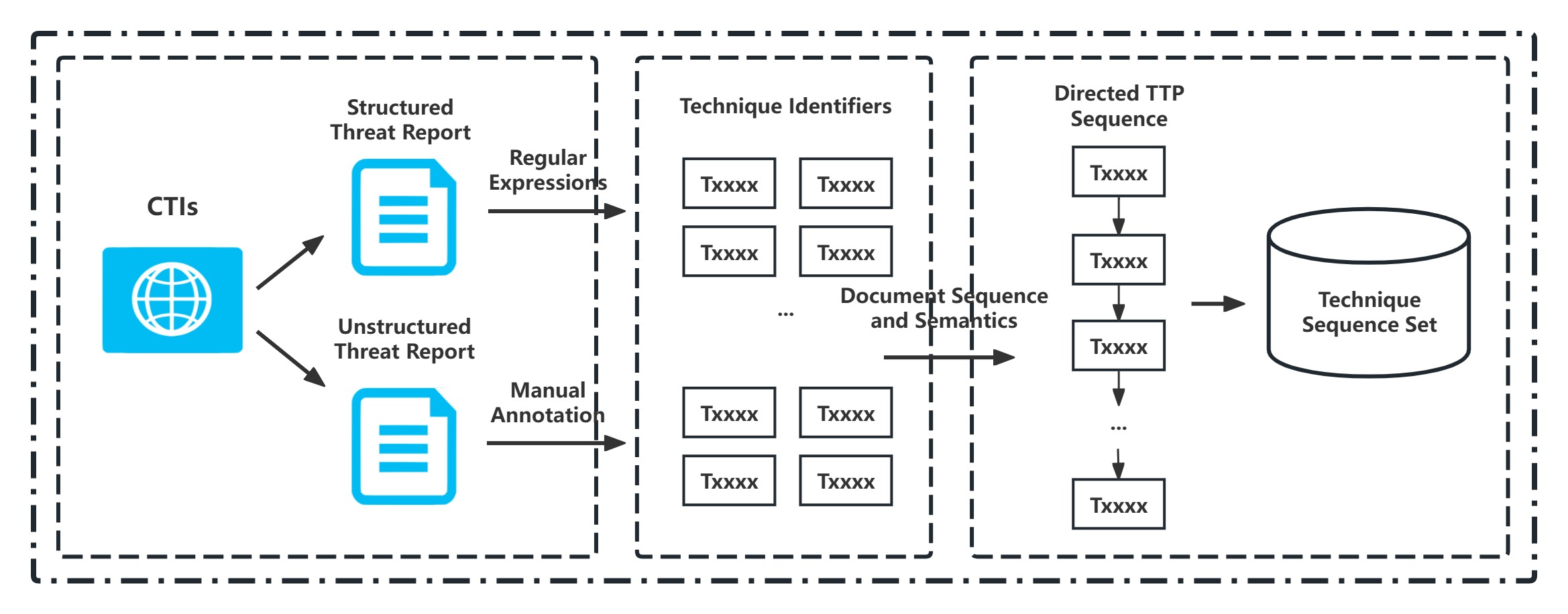} % 替换为你的文件名
  \caption{Construction of the technique sequence set.}
  \label{fig:example4}
\end{figure}

As illustrated in \Cref{fig:example4}, the first stage involves systematically collecting publicly available cyber threat intelligence (CTI) from open-source threat intelligence platforms such as AlienVault OTX, MISP, VirusTotal, and APT28 repositories. The original HTML or PDF reports are downloaded and archived in their raw form. For each threat report, MITRE ATT\&CK technique identifiers (formatted as Txxxx) are directly extracted from the text using regular expressions. In cases where no explicit MITRE IDs are present, we manually inspect the content for specific technique-related terminology and map these terms to the corresponding MITRE ATT\&CK technique identifiers. The extracted techniques are then arranged into a directed TTP sequence according to their order of appearance in the document and the semantic structure of paragraphs or subsections. These sequences are archived in a dedicated technique sequence set for subsequent pattern mining and attribution analysis.

In the second step, we construct the ATT-SPT using the technique sequence set, which comprises a technique pattern tree and a tactic pattern tree—the latter being derived by mapping from the former. For the technique pattern tree (called technique-PT), we extract patterns from the technique sequence set and organize the patterns as a tree structure. Specifically, a root node is first created to initialize the technique-PT. Then, for each attack sequence $ts_{k}$ in the technique sequence set, the tree is expanded as follows:

\begin{itemize}
    \item[•] \textbf{}If the prefix of $ts_{k}$ does not match any branch in the technique pattern tree, $ts_{k}$ will be inserted as a new branch under the root node.
    \item[•] \textbf{}If any prefix of $ts_{k}$ matches the prefix of an existing branch, the unmatched portion of $ts_{k}$ will be appended to the end of the prefix of that branch.
\end{itemize}

\begin{figure*}[t]
  \centering
  \includegraphics[width=\textwidth, height=8cm, keepaspectratio]{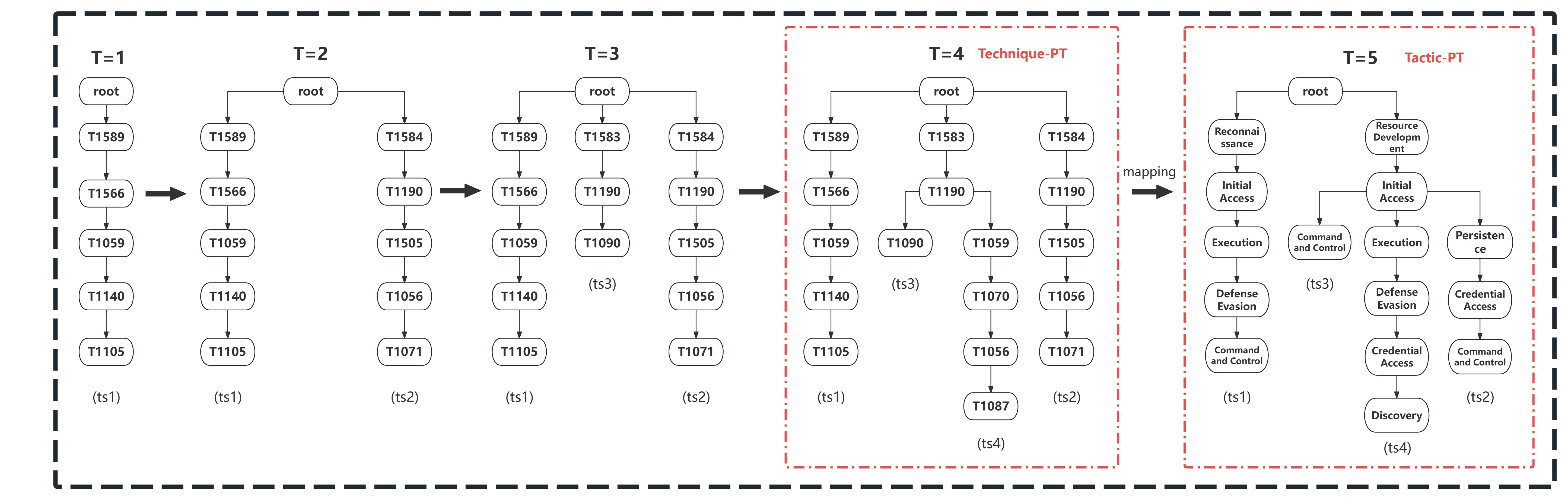} % 替换为你的文件名
  \caption{Construction of ATT-SPT.}
  \label{fig:example5}
\end{figure*}

By iteratively applying this process to each attack technique sequence, the technique pattern tree is incrementally constructed. We give a toy example in \Cref{fig:example5} to illustrate the process. Assume that the technique sequence set contains four attack technique sequences, i.e., $ts_{1}$ = T1589 → T1566 → T1059 → T1140 → T1105, $ts_{2}$ = T1584 → T1190 → T1505 → T1056 → T1071, $ts_{3}$ = T1583 → T1190 → T1090, and $ts_{4}$ = T1583 → T1190 → T1059 → T1070 → T1056 → T1087, the technique pattern tree is constructed through four iterations. Finally, the tactic pattern tree (called tactic-PT) is transferred from the technique-PT by mapping the attack technique of each node into its corresponding attack tactic and filtering out consecutively repeated nodes.

\subsubsection{Edge Threat Evaluation}%3.3.2
In this section, we compute the threat score of each edge in the anomaly subgraphs by referencing the ATT-SPT. Specifically, each directed edge connecting two abnormal nodes represents an attack tactic / technique transition, symbolizing the adversary’s intentions. For example, the technique transition from T1556 to T1570 indicates a shift in the intent from the tactic of credential access to lateral movement. This transition may occur when an attacker successfully locate the desired data within the target system using the stolen credential, and subsequently infiltrates other systems to further compromise. Therefore, we can evaluate the threat by examining whether the attack tactic / technique transition conform to the extracted patterns.

\paragraph{(1) Tactic threat score evaluation}
Given an edge $e_{k}$ representing the attack tactic transition $TA_{s}$→$TA_{t}$ (e.g., Initial Access → Privilege Escalation), this metric is derived by analyzing all reachable paths in the tactic-PT between $TA_{s}$ and $TA_{t}$, accounting for both direct transitions and indirect transitions (transitions within the same tactic are also considered as direct transitions). The scoring strategy weights direct transitions maximally while attenuating indirect transitions proportionally to their transition length, as longer intermediary sequences reduce the coherence of tactic transition. Formally, the tactic threat score $TacticScore$($e_{k}$) is computed as Eq.\eqref{eq:2}, where $P(TA_s,TA_t)$ denotes all paths from $TA_{s}$ to $TA_{t}$ in the tactic-PT. For each path $p \in P(TA_s, TA_t)$, its topological length is quantified as len($p$), while out($TA_{s}$) represents the number of all branch paths starting from tactical $TA_{s}$. Finally, we use a standardized metric parameter $b$ to adjust the discrimination of the scores.

\begin{equation}
    \textit{TacticScore}(e_k) = \frac{\displaystyle\sum_{p \in P(TA_s, TA_t)} \frac{1}{\operatorname{len}(p)}}{\operatorname{out}(TA_s)} \cdot b
    \label{eq:2}
\end{equation}

\paragraph{(2) Technique threat score evaluation}
Technique transition can be viewed as a refinement of tactic transition. While adversaries move across tactical stages, they often select from a variety of techniques to achieve their goals. Within a single tactical stage, multiple techniques can also be orchestrated in tandem to execute complex actions. Unlike tactic transitions that tend to follow higher-level logic, technique transitions are not bound by deterministic rules, allowing attackers considerable flexibility in combining diverse methods to accomplish specific objectives. Therefore, to quantify the threat level of attack technique transitions (i.e., the threat of the edge), we focus on whether the transition exists in attack technique-PT. If a transition path exists between two techniques in the technique-PT, it is considered potentially threatening. Due to that such paths may be either direct or indirect, we assign higher threat scores to direct transitions, lower scores to indirect ones, and minimal scores to transitions not present in the technique-PT. Formally, for a given transition $TA_{s} \to TA_{t}$, the $TechniqueScore$ is computed as Eq.\eqref{eq:3}, where $\Pi_{\mathrm{direct}}$ is an indicator function that is 1 if $TA_{s} \to TA_{t}$ exists in the technique-PT, and 0 otherwise. $P_{i}$ represents all indirect paths from $TA_{s}$ to $TA_{t}$. For each path $p \in P_i$, its topological length is quantified as len(p). In addition, $\varepsilon$ is a negligible probability for unobserved transitions. Finally, we use a standardized metric parameter $a$ to adjust the discrimination of the scores.

\begin{equation}
    \textit{TechniqueScore} = a \cdot \max\left\{ \Pi_{\mathrm{direct}} , \max_{p \in P_{\mathrm{i}}(TA_s, TA_t)} \frac{1}{\mathrm{len}(p)} , \varepsilon \right\}
    \label{eq:3}
\end{equation}

\paragraph{(3) Neighboring interaction evaluation}
In addition to the tactic and technique transitions, the threat score of the edge $e_{k}$ = ($u$, $v$) is also influenced by the interactions with other neighbors. This is reflected in their in-degrees and out-degrees as follows \cite{C16}.

\begin{itemize}
    \item[•] \textbf{}If the in-degree of $u$ is much larger than its out-degree, it indicates that $u$ has been affected by a large number of other nodes and is focused on interacting with $v$.
    \item[•] \textbf{}If the out-degree of $v$ is much larger than its in-degree, it indicates that $v$ is highly dependent on $u$, and has a broad impact on other nodes.
\end{itemize}
Based on these observations, the threat score of the neighboring interactions of $e_{\text{k}}$ (denoted as $ScoreNI(e_k)$) is calculated using Eq.\eqref{eq:4}.
\begin{equation}
    ScoreNI(e_{\text{k}})=\frac{u.indegree}{u.outdegree}+ \frac{v.outdegree}{v.indegree}
    \label{eq:4}  % 设置标签（建议用 eq: 前缀）
\end{equation}
The final threat score of $e_{\text{k}}$ is calculated as the weighted sum of the above mentioned three types, as defined in Eq.\eqref{eq:5}.
\begin{align}
\textit{ScoreE}(e_{\text{k}}) &= \alpha \cdot \textit{TacticScore}(e_{\text{k}}) \nonumber \\
&\quad + \beta \cdot \textit{TechniqueScore}(e_{\text{k}}) \nonumber \\
&\quad + (1-\alpha-\beta) \cdot \textit{ScoreNI}(e_{\text{k}})
\label{eq:5}
\end{align}

As illustrated in Fig.~\ref{fig:example6}(a), we demonstrate the edge scoring mechanism using a tactic-annotated anomaly subgraph and its corresponding tactic pattern tree. The tactic transition score for each edge is computed based on reachable paths in the tactic-PT. For example, the score from Node 3 (Initial Access) to Node 6 (Credential Access) is derived as follows: two paths with lengths 3 and 2 are identified between these tactics in the shown tactic-PT, and with 4 branches originating from Initial Access, Eq.\eqref{eq:2} yields a score of $b \cdot \frac{1/3 + 1/2}{4} = 0.2083b$. This same procedure is applied iteratively to compute the tactic transition scores for all edges in the anomaly subgraph. Critically, the scoring result demonstrates that the tactic transition between Node 4 (Execution) and Node 8 (Execution), belonging to the same tactic, receives the maximum score $b$ as it represents a direct tactic transition. Conversely, no valid path exists from Node 5 (Defense Evasion) to Node 8 (Execution) in the Tactic-PT, resulting in a score of 0. Likewise, the edge from Node 5 (Defense Evasion) to Node 6 (Credential Access) attains a score of $0.5b$. This elevated score reflects both the presence of a direct tactic transition path within the tactic-PT and the low branching factor—only two branches originate from Defense Evasion, one of which is this direct path. This structural feature indicates that, in historical attack cases, transitions from Defense Evasion to Credential Access have occurred with high probability, making this transition more threatening compared to the indirect, multi-hop transition from Node 3 (Initial Access) to Node 6 (Credential Access).

\subsubsection{Attack Path Traversal}%3.3.3
\begin{figure}[t]
  \centering
  \includegraphics[width=\columnwidth]{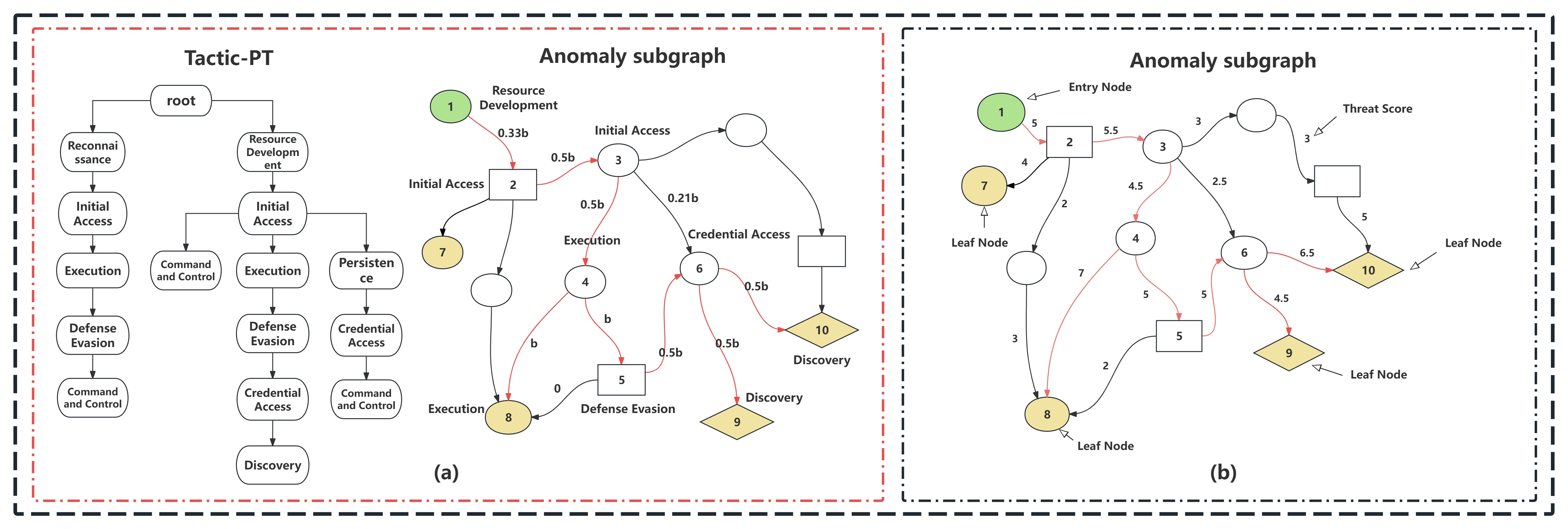} % 替换为你的文件名
  \caption{Attack Path Reasoning.}
  \label{fig:example6}
\end{figure}

After evaluating the threat score of each edge, we try to extract candidate attack paths by traversing the anomaly subgraph. Specifically, TPPR employs a graph traversal algorithm, which backtracks along the reverse edge direction, starting from the node with an out-degree of 0 (i.e., leaf node). At each backtracking step, TPPR selects the edge with the highest threat score, until it reaches a node with an in-degree of 0 (i.e., entry node). The resulting sequence of nodes forms a candidate attack path. Fig.~\ref{fig:example6}(b) gives a toy example, according to the leaf nodes numbered (7)-(10), four candidate attack paths are obtained by tracing forward according to the edge threat weights, i.e., (1)-(2)-(7), (1)-(2)-(3)-(4)-(8), (1)-(2)-(3)-(4)-(5)-(6)-(9), (1)-(2)-(3)-(4)-(5)-(6)-(10). It can be seen that the inferred candidate attack paths cover the real attack path (i.e., (1)-(2)-(3)-(4)-(5)-(6)-(8)-(9)-(10)), but also exist some false positive nodes. Therefore, we introduce the path confidence score to evaluate the candidate attack path obtained by reasoning and retaining the high confidence path to prune potential false positive nodes.

\subsection{Attack Path Scoring and Merging}%3.4
\label{subsec:3.4}
We try to assign a confidence score to each candidate path, retain the high-confidence candidate attack paths, and merge redundant ones into a complete attack path that covers the entire attack chain. Specifically, given a candidate attack path $CP$, we quantify the confidence score of an attack path from the following aspects.

\paragraph{(1) Edge threat}
The assessment of the local threat degree for a $CP$ is performed by computing the average threat score across its edges, according to Eq.\eqref{eq:6}.

\begin{equation}
    EdgeScore(CP)=\frac{\displaystyle\sum_{e_{\text{k}}\in CP.edges}ScoreE(e_{\text{k}})}{EdgeNumber}
    \label{eq:6}  % 设置标签（建议用 eq: 前缀）
\end{equation}
\paragraph{(2) Path length}
This factor is designed to filter out paths that are excessively short yet exhibit a high average threat score, while also mitigating the inflation of scores for spurious attack paths that contain numerous edges, most of which have low threat levels. The path length coefficient is calculated as shown in Eq.\eqref{eq:7}, where the parameter $\lambda$ can be adjusted to optimize sensitivity to path length.

\begin{equation}
    LengthScore(CP)=(1 - e^{-len(CP) / \lambda}) \cdot EdgeScore(CP)
    \label{eq:7}  % 设置标签（建议用 eq: 前缀）
\end{equation}
\paragraph{(3) Anomaly score}
This factor combines the anomaly detection scores of each abnormal node in the candidate path. As TPPR has already detected abnormal nodes and assigned an abnormal score to each node (\Cref{subsec:3.2.1}), we thus average scores of abnormal nodes in the path and retain those with high average scores as potential attack paths.
\begin{equation}
    NodeScore(CP)=\frac{\displaystyle\sum_{v\in Nodes}AbnormalScore(v)}{NodeNumber}
    \label{eq:8}  % 设置标签（建议用 eq: 前缀）
\end{equation}

\begin{figure*}[t]
  \centering
  \includegraphics[width=\textwidth, height=8cm, keepaspectratio]{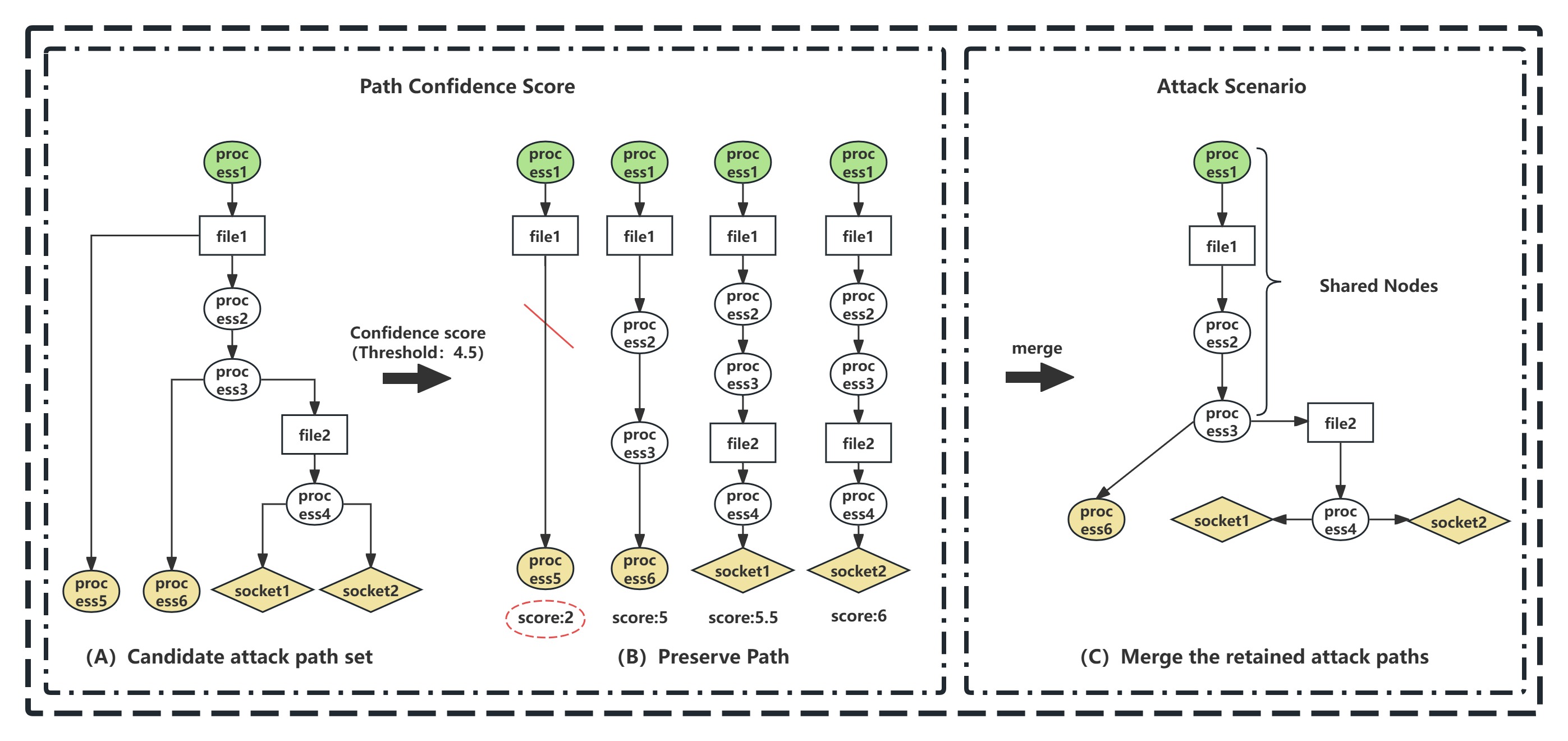} % 替换为你的文件名
  \caption{Merge the retained attack paths to reconstruct attack scenarios.}
  \label{fig:example7}
\end{figure*}

Finally, the attack paths are obtained as follows. First, the confidence score of each candidate attack path $CP$ is calculated as the weighted sum as Eq.\eqref{eq:9}. Second, the candidate attack paths with the confidence score higher than a predefined confidence threshold $\theta$ are retained. Third, according to the node identifiers (such as process ID), the shared nodes between different paths are identified, and overlapping paths are merged into one through these shared nodes. Paths that do not intersect with any others (i.e., those lacking shared nodes) are considered low-confidence and are subsequently filtered out. \Cref{fig:example7} shows an example of high-confidence path retaining and merging. The candidate attack path set comprises four initial paths: $path_1 = p_1 \to f_1 \to p_5$, $path_2 = p_1 \to f_1 \to p_2 \to p_3 \to p_6$, $path_3 = p_1 \to f_1 \to p_2 \to p_3 \to f_2 \to p_4 \to s_1$, and $path_4 = p_1 \to f_1 \to p_2 \to p_3 \to f_2 \to p_4 \to s_2$. Following confidence evaluation, $path_1$ is discarded due to its score of 2, which falls below the predefined confidence threshold. The remaining paths—$path_2$, $path_3$, and $path_4$—share common nodes $p_1$, $f_1$, $p_2$, and $p_3$, with $path_3$ and $path_4$ further sharing $f_2$ and $p_4$. By merging these overlapping nodes, the attack scenario is reconstructed as depicted in Fig.~\ref{fig:example7}(c).

\begin{equation}
\begin{split}
\textit{ConfidenceScore} &= w_{\text{1}} \cdot \textit{LengthScore} \\
&\quad + w_{\text{2}} \cdot \textit{NodeScore}
\end{split}
\label{eq:9}
\end{equation}

\section{Evaluation}
\label{sec:4}
We evaluate the proposed TPPR framework using both public datasets and our own simulated APT attack samples, with the aim of addressing following research questions (RQs):

\begin{itemize}
    \item[$\cdot$] \textbf{RQ1}: How effective is TPPR in reconstructing attack scenarios compared to other state-of-the-art techniques?
    \item[$\cdot$] \textbf{RQ2}: How do various heuristic designs of TPPR influence the reconstruction of attack scenarios?
    \item[$\cdot$] \textbf{RQ3}: How TPPR performs in reconstructing attack scenarios in different attack cases?
    \item[$\cdot$] \textbf{RQ4}: How sensitive is TPPR to parameter configuration?
    \item[$\cdot$] \textbf{RQ5}: How efficient is TPPR in constructing an attack scenario?
\end{itemize}

\subsection{Evaluation Setup}
We deployed Kellect \cite{C13} on a host equipped with Windows 10, CPU i5-10400F, and 16GB of memory to collect audit logs. TPPR was applied to reconstruct the attack scenario from these datasets on a server equipped with Intel Xeon Gold 5218 CPU, 128GB RAM, and Ubuntu 20.04.6.

\subsubsection{Simulated APT attacks}
We simulated five multi-stage APT attacks in the deployed system with background activities including live video stream, general office software usage, gaming and file downloads. The collected audit log data containing more than 100 million events \cite{C13}. Specifically, these attack cases executed in the target system utilized multi-stage tactics and techniques from the ATT\&CK framework. We analyzed these five attack cases, each representing a complete attack campaign as follows:

\textbf{Attack 1 (Phishing Emails):} Adversaries initiate credential compromise via macro-enabled phishing emails. Following successful execution, malicious payloads are deployed to establish persistence through VBScript in startup directories. Attackers conduct reconnaissance using BloodHound, and then perform privilege escalation via ccmstp utility exploitation. Subsequent credential harvesting employs Kerbrute, with defensive evasion through log deletion (TeamViewer artifacts). Final data exfiltration utilizes SMTP transmission.

\textbf{Attack 2 (Account Abuse):} The attacker compromises existing credentials, and then leverages WMI's Invoke-CimMethod to create a hidden scheduled task. They deploy a malicious svchost.exe to System32 for persistence, subsequently using VSS to bypass registry locks and extract SAM/SYSTEM backups for password cracking. Finally, the attacker exfiltrates credentials via encrypted web services.

\textbf{Attack 3 (Content Injection):} The attackers gain initial access by poisoning DNS, HTTP, or SMB responses to deliver malicious scripts. Then, they execute the attrib command to modify system file attributes and bypass permission controls. Subsequently, they conduct domain enumeration using ADSI (Active Directory Service Interfaces), as part of their lateral movement strategy, and ultimately  exfiltrate account data via HTTP.

\textbf{Attack 4 (Content Injection 2):} Adversaries deliver a trojanized executable (e.g., setup.exe) that performs DLL injection into system processes, generates a malicious startup script (startup.cmd) for persistence. After the malicious code is executed, PowerShell is used to delete the generated files, logs and registries to evade detection by security tools.

\textbf{Attack 5 (Service suspension):} The attackers deploy malicious USB devices with firmware-level implants to gain initial access. Then, they leverage Windows management tools (SchTasks/WMI) to establish persistence via scheduled tasks, disable critical security services via command-line, and ultimately deploy AES encryption for system disruption.

\subsubsection{Public datasets}
In addition, we evaluated the effectiveness of our method on public DARPA datasets , which is a mature benchmark for threat detection research and contains  data collected by three different mechanisms: CADETS, THEIA, and CLEARSCOPE. Simultaneously, unlike real-world enterprise logs, the DARPA dataset has lower noise interference and more sparse distribution of malicious nodes \cite{C46}. We will describe the test cases of the DARPA dataset in detail in the Case Study section.

\begin{table}[htbp]
  \centering
  \caption{Statistics of anomaly subgraphs generated by all 10 attacks.}
  \label{tab:tb2}  % 标签名
  \begin{tabular}{llll}
    \toprule
    \textbf{Attack Cases} & \textbf{\#Vertice} & \textbf{\#Edge} & \textbf{\#CE}\\
    \midrule
    Phishing Emails & 291 & 867 & 293 \\
    Account Abuse & 155 & 572 & 164 \\
    Content Injection & 216 & 788 & 219 \\
    Content Injection 2 & 207 & 603 & 207 \\
    Service suspension & 160 & 526 & 158 \\
    Five Dir Case 1 & 327 & 618 & 530 \\
    Five Dir Case 3 & 398 & 1062 & 631 \\
    Theia Case 1 & 2287 & 2663 & 2609 \\
    Theia Case 3 & 4507 & 13836 & 8538 \\
    Theia Case 5 & 338 & 771 & 458 \\
    \midrule
    AVG & 888.60 & 2230.60 & 1380.70 \\
    \bottomrule
  \end{tabular}
\end{table}

\subsubsection{Obtaining Ground Truth of Attacks}
For each attack case, we conduct an in-depth manual analysis to extract the complete attack chain, enabling a comprehensive evaluation of the attack reconstruction. In specific, we perform node-level anomaly detection within collected system events. Abnormal nodes are retained, and the graph are pruned to only include critical events, thereby obtaining a refined anomaly subgraph focused on attack-relevant behavior. \Cref{tab:tb2} presents the statistics of the anomaly subgraphs generated for these 5 attacks and 5 host attacks contained in the DARPA dataset(rows 7 to 11) \cite{C14}. The ”Attack Cases” column lists the attack cases. The "\#V" and "\#E" columns represent the number of abnormal nodes and their related nodes in the anomaly subgraph before pruning the event edges, as well as the corresponding number of event edges, while the “\#CE” column indicates the number of critical edges remaining after graph simplification. As the results indicate, the original anomaly subgraphs contained a substantial number of edges unrelated to the actual attack. The simplification operation significantly reduced graph complexity thereby improving the efficiency and clarity of subsequent path reasoning.

\begin{table*}[t]
\centering
\caption{Attack scenarios reconstructed by each approach.}
\label{tab:tb3}
\begin{tabular}{lccccccccc}
\toprule
\multirow{2}{*}{Attack Cases} & \multicolumn{3}{c}{TPPR} & \multicolumn{3}{c}{SPARSE} & \multicolumn{3}{c}{DepImpact} \\
\cmidrule(lr){2-4} \cmidrule(lr){5-7} \cmidrule(lr){8-10}
 & Pre. & Recall & F1 & Pre. & Recall & F1 & Pre. & Recall & F1 \\
\midrule
Phishing Emails    & 84.21\% & 100\%   & 91.43\% & 28.57\% & 25.00\% & 26.67\% & 50.00\% & 80.00\% & 61.54\% \\
Account Abuse      & 78.57\% & 91.67\% & 84.62\% & 40.00\% & 33.33\% & 36.36\% & 66.67\% & 80.00\% & 72.73\% \\
Content Injection  & 75.00\% & 81.82\% & 78.26\% & 37.50\% & 27.27\% & 31.58\% & 20.00\% & 66.67\% & 30.77\% \\
Content Injection 2 & 86.67\% & 76.47\% & 81.25\% & 38.89\% & 41.18\% & 40.00\% & 11.11\% & 50.00\% & 18.18\% \\
Service suspension & 69.23\% & 90.00\% & 78.26\% & 37.50\% & 30.00\% & 33.33\% & 58.33\% & 70.00\% & 63.64\% \\
Five Dir Case 1    & 37.50\% & 100.00\% & 54.55\% & 42.86\% & 100.00\% & 60.00\% & 33.33\% & 100.00\% & 50.00\% \\
Five Dir Case 3    & 35.29\% & 100.00\% & 52.17\% & 27.27\% & 100.00\% & 42.85\% & 20.69\% & 100.00\% & 34.29\% \\
Theia Case 1       & 25.81\% & 100.00\% & 41.03\% & 18.42\% & 87.50\% & 30.43\% & 12.31\% & 100.00\% & 21.92\% \\
Theia Case 3       & 22.12\% & 89.29\% & 35.45\% & 33.87\% & 75.00\% & 46.67\% & 24.35\% & 100.00\% & 39.16\% \\
Trace Case 5       & 30.77\% & 80.00\% & 44.45\% & 29.41\% & 100.00\% & 45.45\% & 27.78\% & 100.00\% & 43.48\% \\
\midrule
AVG & 54.52\% & 90.93\% & 64.15\% & 33.43\% & 61.93\% & 39.33\% & 32.46\% & 84.67\% & 43.57\% \\
\bottomrule
\end{tabular}
\end{table*}
\textbf{Evaluation Metrics.} We define evaluation metrics as: True Positive (TP) = correctly identified attack nodes, False Negative (FN) = missed attack nodes, False Positive (FP) = falsely reported benign nodes. The derived metrics include Precision ($\mathrm{TP}/(\mathrm{TP}+\mathrm{FP})$), Recall ($\mathrm{TP}/(\mathrm{TP}+\mathrm{FN})$), F1-score ($2\cdot\mathrm{Precision}\cdot\mathrm{Recall}/(\mathrm{Precision}+\mathrm{Recall})$), and False Negative Rate, FNR ($\mathrm{FN}/(\mathrm{TP}+\mathrm{FN})$). These metrics collectively measure attack reconstruction accuracy.

\subsection{How effective is TPPR in reconstructing attack scenarios compared to other state-of-the-art techniques?}

To demonstrate the effectiveness of TPPR in reconstructing attack scenarios, we compared TPPR with state-of-the-art methods SPARSE \cite{C15} and DepImpact \cite{C16} as follows:
\begin{itemize}
    \item[•] \textbf{}SPARSE employs a hybrid approach integrating real-time state propagation and path-level analysis. It constructs a Suspicious Semantic Graph (SSG) via automated label propagation based on lightweight semantic rules, from which suspicious flow paths are extracted to form a Critical Component Graph (CCG). In contrast to TPPR, which delves into the extraction of attack behavior patterns, SPARSE prioritizes efficient evidence filtering. It rapidly identifies key event sequences relevant to the POI without interpreting the attack strategies.
    \item[•] \textbf{}DepImpact operates as a graph filtering technique based on multi-feature modeling and influence propagation. It computes edge weights using system features and propagates dependency influence backward from the POI to identify critical paths. In contrast to TPPR's focus on reconstructing semantically meaningful attack scenarios with TTP context, DepImpact adopts a data-feature-driven approach. It prioritizes the identification of POI-related event sequences through quantitative feature evaluation, focusing on system-level causality analysis rather than interpreting global attack intent.
\end{itemize}

Since both SPARSE and DepImpact require POI nodes as starting points, we annotate POI nodes using our anomaly detection model. From the comparison results in \Cref{tab:tb3}, TPPR outperforms across all metrics, with more attack nodes recovered and less noises introduced. In particular, TPPR’s F1 score (64.15\%) is much higher than that of SPARSE (39.33\%) and DepImpact (43.57\%), demonstrating robustness of recovering APT attack chains with high precision.

\begin{table*}[t] % table*环境实现跨双栏
\centering
\caption{Effects of TPPR and its variants on attack scenario reconstruction.}
\label{tab:tb4}
\footnotesize % 缩小字体以适应宽度
\begin{tabular}{
  @{}>{\raggedright\arraybackslash}p{2.5cm} % 第一列固定宽度
  *{4}{S[table-format=2.2]S[table-format=3.2]S[table-format=2.2]} % 每组3列数据
}
\toprule
\multirow{2}{*}{\makecell{Attack Cases}} & 
\multicolumn{3}{c}{TPPR} & 
\multicolumn{3}{c}{\makecell{w/o TTP}} & 
\multicolumn{3}{c}{\makecell{w/o PS}} & 
\multicolumn{3}{c}{\makecell{w/o ATT-SPT}} \\
\cmidrule(lr){2-4} \cmidrule(lr){5-7} \cmidrule(lr){8-10} \cmidrule(lr){11-13}
 & \multicolumn{1}{c}{Pre.} & \multicolumn{1}{c}{Recall} & \multicolumn{1}{c}{F1} 
 & \multicolumn{1}{c}{Pre.} & \multicolumn{1}{c}{Recall} & \multicolumn{1}{c}{F1}
 & \multicolumn{1}{c}{Pre.} & \multicolumn{1}{c}{Recall} & \multicolumn{1}{c}{F1}
 & \multicolumn{1}{c}{Pre.} & \multicolumn{1}{c}{Recall} & \multicolumn{1}{c}{F1} \\
\midrule
Phishing Emails    & 84.21\% & 100.00\% & 91.43\% & 42.86\% & 18.75\% & 26.09\% & 10.46\% & 100.00\% & 18.93\% & 42.86\% & 18.75\% & 26.09\% \\
Account Abuse      & 78.57\% & 91.67\% & 84.62\% & 25.00\% & 22.22\% & 23.53\% & 7.79\% & 100.00\% & 14.46\% & 62.50\% & 55.56\% & 58.82\% \\
Content Injection  & 75.00\% & 81.82\% & 78.26\% & 30.77\% & 36.36\% & 33.33\% & 9.02\% & 100.00\% & 16.54\% & 45.45\% & 50.00\% & 47.62\% \\
Content Injection 2 & 86.67\% & 76.47\% & 81.25\% & 11.11\% & 5.88\% & 7.69\% & 8.29\% & 100.00\% & 15.32\% & 22.22\% & 11.76\% & 15.38\% \\
Service suspension & 69.23\% & 90.00\% & 78.26\% & 28.57\% & 18.18\% & 22.22\% & 18.03\% & 100.00\% & 30.56\% & 46.15\% & 60.00\% & 52.17\% \\
\midrule
AVG & 78.74\% & 87.99\% & 82.76\% & 27.66\% & 20.28\% & 22.57\% & 10.72\% & 100.00\% & 19.16\% & 43.84\% & 39.21\% & 40.02\% \\
\bottomrule
\end{tabular}
\end{table*}

\begin{figure*}[t]
  \centering
  \includegraphics[width=\textwidth, height=8cm, keepaspectratio]{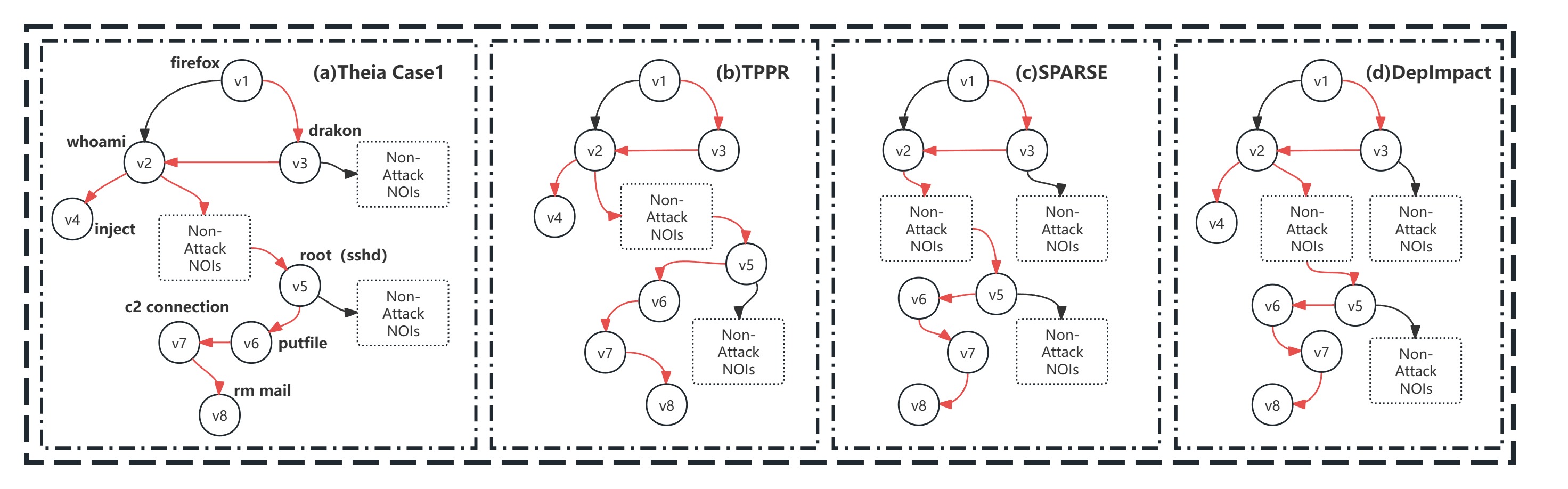} % 替换为你的文件名
  \caption{Crucial provenance graph analysis of Case \#1.}
  \label{fig:example8}
\end{figure*}
\begin{figure*}[t]
  \centering
  \includegraphics[width=\textwidth, height=10cm, keepaspectratio]{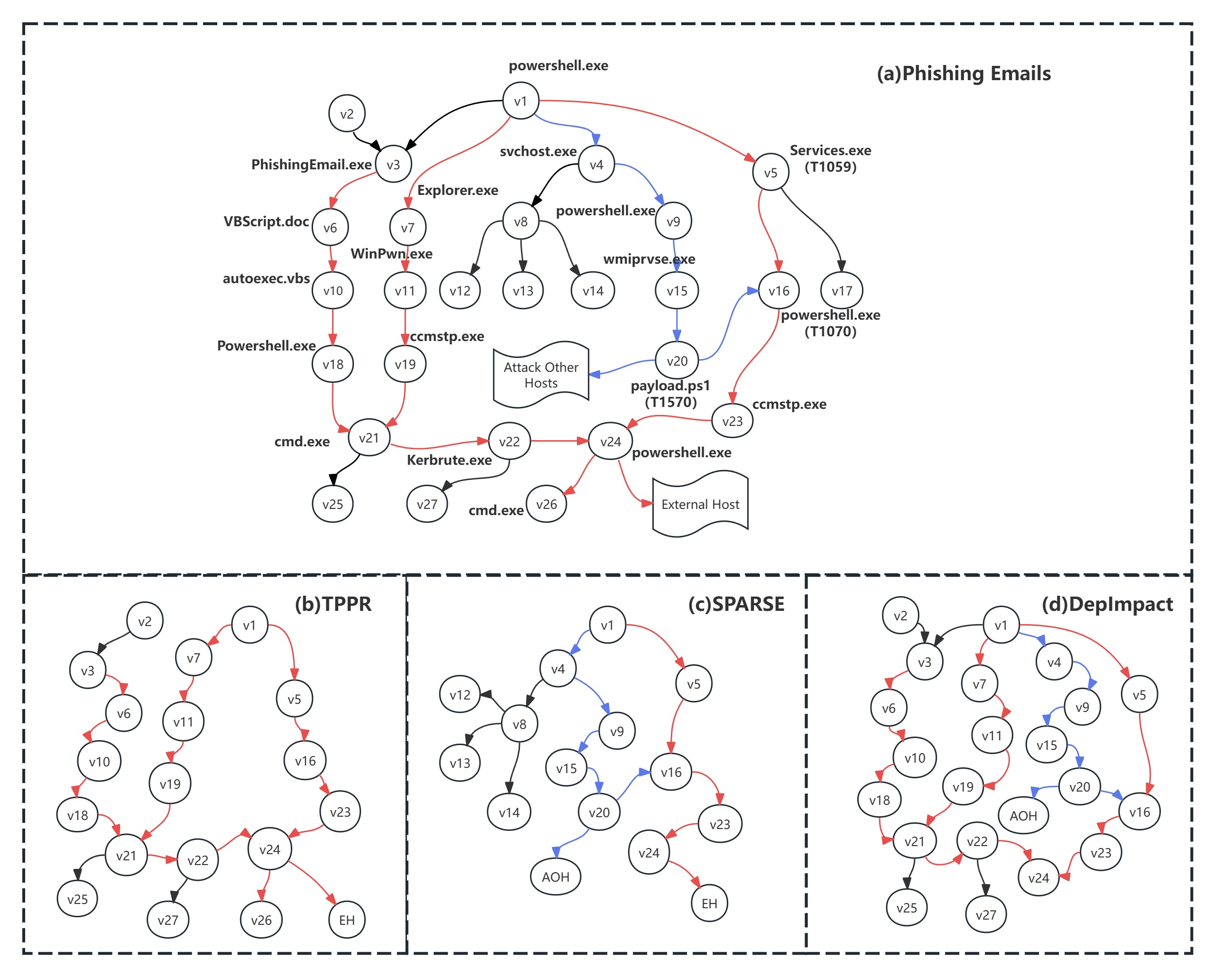} % 替换为你的文件名
  \caption{Crucial provenance graph analysis of Case \#2.}
  \label{fig:example9}
\end{figure*}

We summarize these performance gaps for the following reasons: First, SPARSE extracts suspicious flow paths based on semantic and temporal relationships from a given POI node, which struggles to filter out benign nodes that are adjacent to the attack. This can lead to the inclusion of extraneous nodes, reducing the precision of the reconstructed attack scenario. In addition, long and complex attack chains cannot be fully recovered by solely employing suspicious path extraction based on semantic and temporal relationships. Therefore, the attack scenario graph generated by SPARSE contains a large number of FPs (141 out of 204), resulting in poor performance (F1=39.33\%).

Second, DepImpact performs well in Recall (84.67\%), but poorly in precision (32.46\%), which indicates a tendency to generate a large volume of FPs. The primary reason for this issue lies in DepImpact’s heuristics of selecting the overlapping subgraph between the backward and forward dependency graphs as the output. Although it is effective in covering multi-stage attacks, a large number of irrelevant nodes are inadvertently included, thus reducing precision. 

Finally, the best-performing TPPR achieves an impressive F1=64.15\%, we attribute its success to two main reasons: (1) By incorporating TTP patterns, TPPR effectively filters out nodes that do not belong to the same attack campaign, thereby enhancing precision. (2) It computes a confidence score for paths by jointly considering local edge semantics and global path semantics, allowing for effective differentiation between paths from different attack campaigns.

\subsection{How do various heuristic designs of TPPR influence the reconstruction of attack scenarios?}
We evaluate the impact of different modules of TPPR in attack scenario reconstruction. Specifically, we compare TPPR with three variants: 
\begin{itemize}
    \item \textbf{w/o TTP annotation}: 
    It refers to the variant without TTP annotation. Specifically, it randomly assigns TTP types to nodes in the anomaly subgraph.
    
    \item \textbf{w/o PS}: 
    It refers to the variant without path scoring. Specifically, it does not perform confidence assessment on the candidate attack paths obtained by path reasoning, but directly merge all attack paths obtained by path reasoning as the reconstructed attack scenario.
    
    \item \textbf{w/o ATT-SPT}: 
    It refers to the variant without ATT-SPT. Specifically, it replaces the ATT-SPT in the knowledge base with a simple attack sequence set.
\end{itemize}

The results in \Cref{tab:tb4} highlight the superior performance of TPPR which significantly surpasses the variants by 72.73\%, 76.85\% and 51.64\% in F1 score, respectively. This significant performance difference validates the core role of node-level TTP recognition in TPPR’s path reasoning. Further analysis shows that although the candidate attack path reconstructed without path confidence evaluation achieved a perfect Recall (100\%), its average precision was only 10.72\%. This is because the algorithm retains all potential attack paths based on semantic features and threat propagation weights, resulting in the generated scenario containing a large number of irrelevant nodes. This observation highlights the critical role of path-level confidence evaluation in filtering out irrelevant paths and improving precision.  Meanwhile, a significant drop in F1-score was seen in the w/o ATT-SPT variant, which confirms the importance of the edge scoring mechanism based on ATT-SPT for attack path reasoning.

\subsection{Case Study}
To answer RQ3, we conduct a granular analysis of individual attack instances. The aforementioned experiments demonstrate that TPPR significantly outperforms the baseline methods in reducing both false positive and false negative during attack scenario reconstruction. Most of the reasoning errors stem from the inherent complexity of the attack scenarios including: (a) multiple interdependent attack chains with complex coupling relationships and (b) a multi-stage attack sequence traversing various defense perimeters. The reason why baseline methods cannot effectively address these issues is that they rely solely on local semantic or contextual node information, which fundamentally lack the capability to discern global attack intention patterns. As a case in point, when confronted with intersecting attack paths sharing common nodes, conventional techniques like SPARSE and DepImpact inherently aggregate these paths into a single composite scenario due to their reliance on static topological features. In contrast, TPPR can distinguish different attack trajectories that follow different TTP sequential patterns, effectively decoupling intertwined paths into their constituent attack scenarios. We empirically validate this capability through the following experimental case analysis.

\textbf{Case \#1:} The “Theia Case 1” in the DARPA TC dataset captures a client-side APT attack, characterized by the exploitation of a Firefox browser vulnerability to deploy a backdoor implant, which subsequently establishes a persistent reverse shell to a remote command-and-control (C2) server.

For this case, TPPR, SPARSE, and DepImpact all demonstrated robust performance. They exhibited notably high attack chain coverage rates, effectively capturing the attack progression.

Analysis of Theia Case 1 (\Cref{fig:example8}) reveals two critical characteristics of the evaluated attack scenario: 1) The attack predominantly follows a linear, single-chain dependency structure. 2) Genuine attack nodes are sparsely distributed amidst a high volume of non-attack abnormal nodes. In other words, most provenance graphs in the DARPA dataset contain only a single attack campaign, so what the models need to do is simply connecting malicious nodes in chronological order while filtering out benign nodes. Under this condition, all three methods achieved complete attack scenario reconstruction with comparable performance. Nevertheless, TPPR still demonstrates superior false positive suppression, reconstructing more concise attack scenarios compared to baseline methods.

\textbf{Case \#2:} The evaluated attack scenario, “Phishing Emails”, is derived from our simulated APT attacks. It initiates credential compromise via phishing emails, followed by persistence establishment, privilege escalation, and the systematic collection and exfiltration of sensitive data. Notably, the scenario incorporates a process-masquerading chain that embeds malicious execution within legitimate system process sequences to complicate forensic analysis.

The assessment of the attack scenario reconstruction in this case reveals a considerable performance edge for TPPR over SPARSE and DepImpact in terms of precision and recall, highlighting its enhanced capability in analyzing coupled attack chains.

In a real network environment, multiple attack campaigns may occur simultaneously, and they may even share some common nodes (e.g., common processes). As illustrated in \Cref{fig:example9}, the anomaly subgraph constructed from system logs of phishing attack cases contains two attack paths: the red path represents the actual attack chain of the phishing campaign, while the blue path represents another attack chain, namely the process-masquerading chain of malicious behavior. These two attack chains intersect at the critical process node $v_{16}$ (tagged with technique T1070), which serves as a shared node reused by different attack chains to achieve distinct adversarial objectives. The TPPR method employs TTP sequential patterns to guide path reasoning. When backtracking reaches node $v_{16}$, two predecessor nodes are detected: $v_{5}$ (tagged with T1059) and $v_{20}$ (tagged with T1570). Based on ATT-SPT correlation analysis, since the transition probability from T1059 → T1070 is significantly higher than that of T1570 → T1070, the path inference process prioritizes $v_{5}$ while disregarding $v_{20}$, thereby effectively mitigating the attack chain coupling problem that could otherwise lead to scenario ambiguity.

In contrast, the SPARSE method focuses primarily on semantic relationships, relying on POI quantities and their temporal dependencies for path inference. For instance, when reasoning backward from an external host node to $v_{24}$, the temporal precedence of $v_{23} \to v_{24}$ over $v_{22} \to v_{24}$ causes the overlooking of the left-side critical attack chain. However, if multiple correct POIs are identified, SPARSE can successfully reconstruct this attack path, indicating that it is more suitable for single-chain attack scenarios.

Meanwhile, DepImpact leverages forward and backward causal analysis between POIs and entry nodes, which allows for a more comprehensive coverage of multiple attack chains. However, in cases of coupled attack chains, it struggles to discern the distinct attack intents of distinct attack paths, thereby limiting its effectiveness in multi-chains attack scenarios.

\begin{table*}[t] % table*环境实现跨双栏
\centering
\caption{Runtime performance of TPPR and baseline approach.}
\label{tab:tb5}
\small
\begin{tabular}{
  @{}l
  S[table-format=3.2]
  S[table-format=1.2]
  S[table-format=1.2]
  S[table-format=1.3]
  S[table-format=1.2]
  S[table-format=1.2]
  @{}
}
\toprule
\multirow{2}{*}{Attack Cases} & 
\multicolumn{1}{c}{\multirow{2}{*}{\makecell{Anomaly Subgraph Min.(s)}}} & 
\multicolumn{2}{c}{Weight Computation(s)} & 
\multicolumn{3}{c}{Attack Scenario Recon.(s)} \\
\cmidrule(lr){3-4} \cmidrule(lr){5-7}
 & & {TPPR} & {DepImpact} & {TPPR} & {SPARSE} & {DepImpact} \\
\midrule
Phishing Emails    & 234.82 & 2.60 & 0.82 & 0.003 & 0.06 & 1.35 \\
Account Abuse      & 137.45 & 2.43 & 0.96 & 0.002 & 0.05 & 1.03 \\
Content Injection  & 134.61 & 1.77 & 0.44 & 0.001 & 0.05 & 0.86 \\
Content Injection 2 & 123.97 & 2.71 & 0.94 & 0.003 & 0.06 & 1.12 \\
Service suspension & 117.74 & 2.23 & 0.91 & 0.002 & 0.06 & 0.99 \\
\midrule
AVG & 149.72 & 2.35 & 0.81 & 0.0022 & 0.056 & 1.07 \\
\bottomrule
\end{tabular}
\end{table*}

\begin{figure*}[t] % 使用figure*跨双栏
\centering
\begin{minipage}[t]{0.48\textwidth}
\centering
\includegraphics[width=\linewidth]{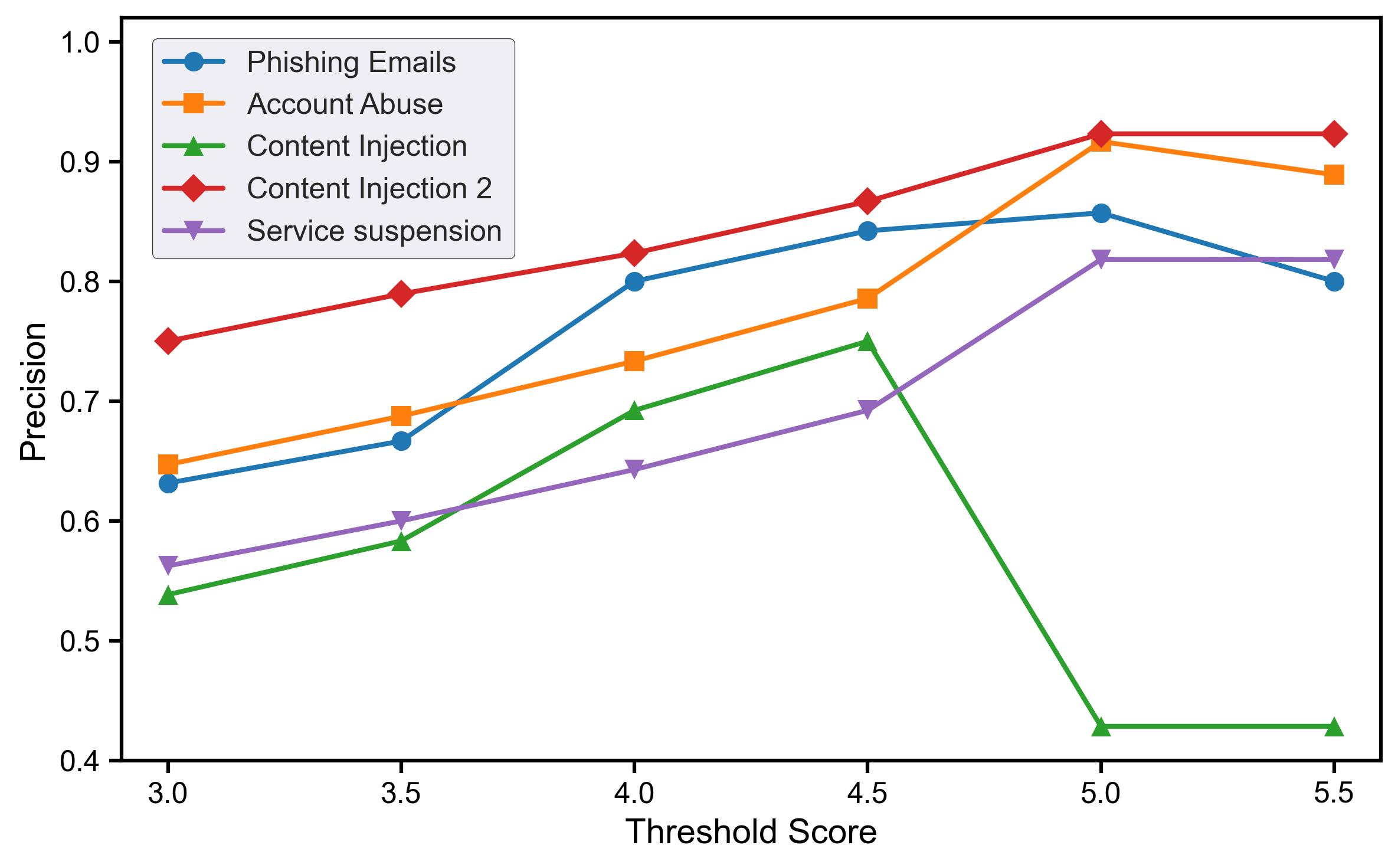} % 第一张图
\caption{Precision using different $\theta$.}
\label{fig:example10}
\end{minipage}
\hfill
\begin{minipage}[t]{0.48\textwidth}
\centering
\includegraphics[width=\linewidth]{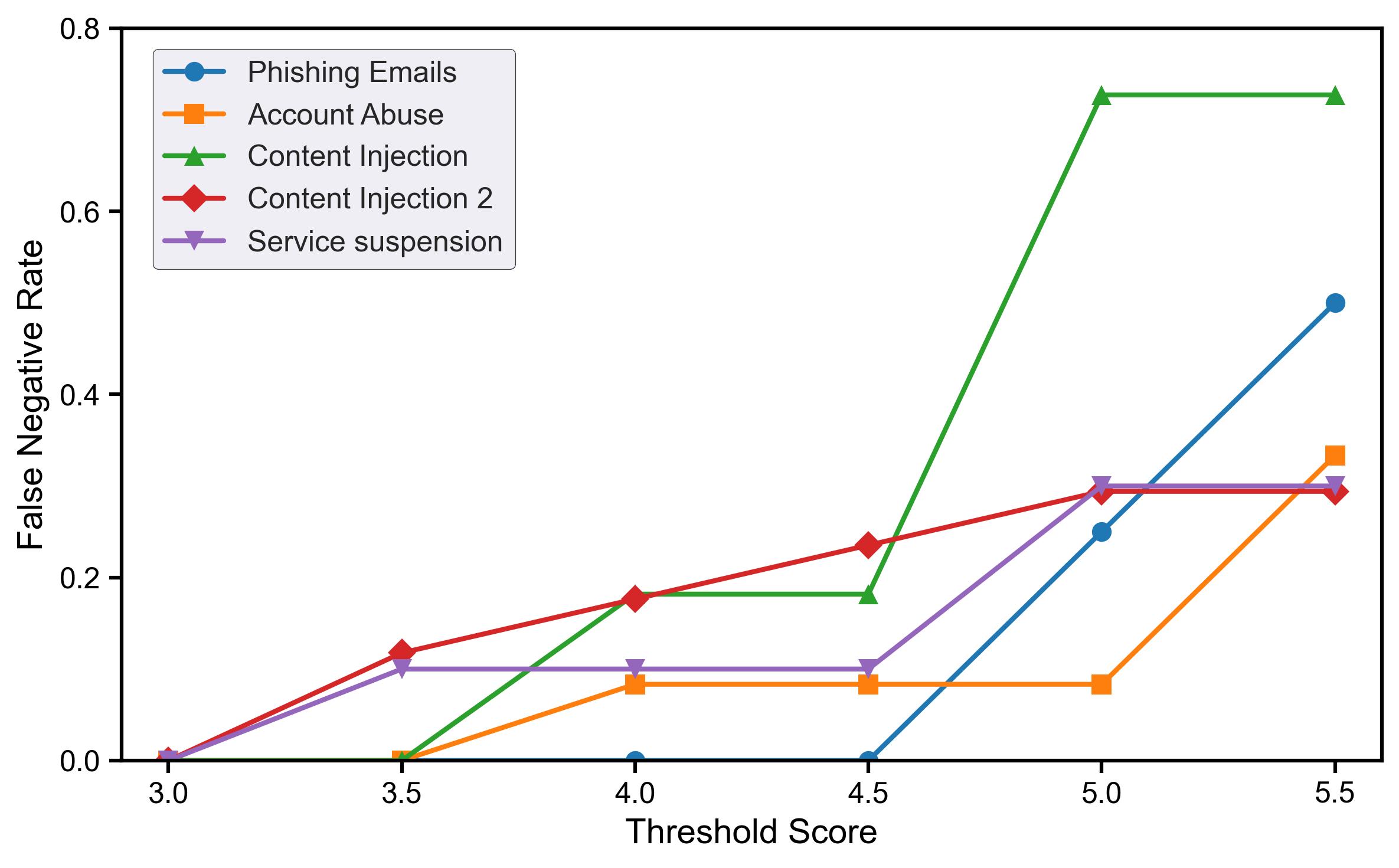} % 第二张图
\caption{False Negative Rate using different $\theta$.}
\label{fig:example11}
\end{minipage}
\end{figure*}
\subsection{How sensitive is TPPR to parameter configuration?}
The confidence threshold $\theta$ (\Cref{subsec:3.4}) plays a critical role in balancing detection sensitivity and precision: lower values retain more attack-relevant nodes but introduce noise, while higher values yield cleaner paths at the risk of omitting critical attack steps. To assess this trade-off, we evaluated TPPR’s performance under varying threshold settings, focusing on changes in both Precision and FNR.

As illustrated in \Cref{fig:example10} and \Cref{fig:example11}, employing a lower $\theta$(3.0-3.5) results in a higher number of candidate inference paths in attack scenarios. While this configuration achieves a near-zero FNR, the Precision remains suboptimal. As the $\theta$ is progressively elevated, Precision exhibits a gradual ascent to its peak; however, this coincides with a dramatic surge in the FNR. This phenomenon stems from the over-pruning of candidate inference paths, producing an excessively streamlined scenario that omits genuine attack nodes. The proposed TPPR framework exhibits strong adaptability to variations in the threshold $\theta$, consistently achieving accurate attack scenario reconstruction across diverse attack cases within a defined $\theta$ range. This demonstrates TPPR’s notable generalization capability and operational stability.

\subsection{How efficient is TPPR in constructing an attack scenario?}
To rigorously evaluate the performance of TPPR, we systematically measured the execution time of its critical components across the entire processing pipeline. As shown in \Cref{tab:tb5}. TPPR completes an attack scenario reconstruction—comprising anomaly subgraph mining, threat-weighted edge computation, and attack scenario reconstruction—in 152.07 seconds on average. Comprehensive comparative analysis with SPARSE and DepImpact reveals: (1) TPPR’s ATT-SPT-based threat weight modeling (2.35 sec) is inferior to DepImpact’s dependency propagation execution time (0.81 sec) due to its tactic and technique transition probability framework that calculates edge weights through TTP correlation, trading temporal efficiency for 67.9\% reconstruction precision gains. (2) During attack graph reconstruction, TPPR achieves 0.0022-second reconstruction runtime—outperforming SPARSE (0.056 sec) by 25× and DepImpact (1.07 sec) by 486×—through its path confidence scoring mechanism that prunes irrelevant attack behaviors via confidence thresholds. Although TPPR incurs a slight increase in runtime during initial phases, its TTP-based threat scoring model ultimately achieved exceptional attack scenario fidelity, with an attack scenario reconstruction accuracy of 54.52\% and a recall rate of 90.93\%.

\section{Discussion}
\paragraph{\textbf{Practical Utility}} 
Attack provenance aims to reconstruct the attack scenario graph that encapsulates the sequence and relationships of malicious activities. As a comprehensive attack investigation framework, TPPR integrates a variety of advanced techniques to enhance the reconstruction of complex, multi-stage attack chains. By combining node-level anomaly detection with TTP recognition mechanisms \cite{C19}, TPPR effectively filters out a large number of attack-irrelevant nodes and edges, and annotates abnormal nodes with their corresponding technique or tactic defined in the ATT\&CK framework. These synergetic components enable TPPR to perform rapid and efficient reasoning along potential attack paths, revealing high-priority attack behaviors throughout the entire attack cycle. By concentrating on the most relevant elements within the attack scenario graph, TPPR provides actionable insights for threat mitigation and proactive defense strategies.

\paragraph{\textbf{Generalization to Real-World Attacks}} 
APTs are characterized by their multi-stage nature, employing a coordinated set of attack tactics and techniques to compromise target systems. These tactics and techniques often exhibit intrinsic correlations, as defined by the MITRE ATT\&CK framework. TPPR combines historical attack behavior insights with the latest adversarial strategies to extract the correlations between tactics and techniques, facilitating the reconstruction of ATT-SPT. In addition, the TPPR continuously evolves by incorporating novel attack patterns derived from emerging threat vectors, which enables iterative refinement and expansion of sequential patterns, thereby improving the adaptability to previously unseen attack scenarios.

\paragraph{\textbf{Limitations}} 
While TPPR has made significant progress in attack scenario reconstruction through the integration of external TTP knowledge, its effectiveness is inherently dependent on the performance of the underlying anomaly detection module: (1) the initial anomaly subgraph construction is constrained by detection recall limitations;  (2) missing critical attack nodes disrupts the continuity of reconstructed attack chains. These limitations are prevalent in most behavior-based detection systems. Notably, TPPR’s hierarchical analysis framework associates discrete abnormal nodes with known TTP sequential patterns, effectively mitigating these challenges, as demonstrated in our evaluation in \Cref{sec:4}. Such modular design is well suited for future enhancements, such as incorporating incremental learning techniques for ATT-SPT updates and integrating advanced node-level detection techniques.

\section{Related Work}
\paragraph{\textbf{Provenance-Based Attack Investigation}} 
Provenance-based attack investigation serves as a critical method for reconstructing APT attack scenarios. In dependency graph optimization, NoDoze \cite{C22} employs execution profile-driven automated triage to significantly reduce graph scale while preserving critical attack paths, thereby mitigating security analyst alert fatigue; complementarily, PrioTracker \cite{C23} utilizes dynamic priority scheduling based on node fanout to reduce manual edge through high-risk path prioritization, though it risks misclassifying high-degree nodes in complex software ecosystems (e.g., browsers). For distributed environments, cross-host provenance tracing enables comprehensive reconstruction of APT lateral movement chains via correlation of multi-source audit logs (e.g., system calls, NetFlow records), exemplified by Gao et al.'s information flow tagging \cite{C5} and Nan et al.'s cross-host causal inference \cite{C30} resolving attack path fragmentation in cloud-native infrastructures. CTI-augmented methodologies have gained significant traction: studies \cite{C31,C32,C33} demonstrate real-time detection of stealthy attacks through proactive threat hunting query matching against provenance graphs, while Milajerdi et al. \cite{C4} and Zhao et al. \cite{C35} leverage structured threat intelligence for semantic node annotation \cite{C36}, with ATT\&CK framework alignment enhancing attack intent recognition accuracy. Persistent limitations include dependency explosion from backward causality analysis initiating at POIs \cite{C24} and semantic ambiguity wherein conventional pruning heuristics (e.g., temporal/data-flow features) inadvertently eliminate legitimate attack edges due to contextual misinterpretation \cite{C9}. TPPR achieves more accurate attack investigations by combining fine-grained provenance-based analysis with external threat intelligence. It can also work with other approaches. In the pruning phase, Template-Based Graph Compression can be used to merge similar nodes \cite{C25}; low-value log events can also be eliminated through the Log Garbage Collection mechanism \cite{C26} to reduce noise interference.
\paragraph{\textbf{Anomaly Detection}} 
Anomaly detection has evolved substantially. The early systems depended on threshold-driven statistical heuristics and signature matching \cite{C37,C38}. ML/DL-based anomaly detection minimizes manual intervention. Among these: Supervised learning and ensemble methods leverage labeled data for model training but require high-fidelity annotations \cite{C39,C40}. Unsupervised learning techniques can effectively identify statistical anomalies in log sequences but are constrained by their semantic-agnostic nature, preventing meaningful interpretation of event relationships \cite{C4,C41,C42}. Graph neural networks (GNNs) enhance detection accuracy via provenance graph analysis but incur significant computational overhead \cite{C3,C34}. APTSHIELD framework \cite{C43} introduces a semantically-rich attack chain modeling framework, abstracting multi-stage adversarial semantics into interpretable semantic transitions for improved threat attribution. However, current methodologies remain primarily focused on node-level threat detection, with limited exploration of cross-domain threat causality in complex attack scenarios. TPPR innovatively leverages insights from node-level detection approaches \cite{C8,C19,C27,C28,C29} and fuses them with CTI-enhanced to enable high-fidelity reconstruction of multi-stage attack scenarios while minimizing false positives.

\section{Conclusion}
We proposed a framework called TPPR, which generates a provenance graph from system logs through causal analysis, and performs node-level anomaly detection and TTP type recognition on it, retaining abnormal nodes and their key event edges, and annotating the TTP types of the nodes. Further, the event edges of the anomaly subgraph are scored for threat weight by mining the ATT-SPT and node neighboring interaction relationships, and candidate attack paths are obtained by walking according to the threat weight in path reasoning. TPPR evaluates the confidence of the path through multiple features of the path, retains and merges high-confidence paths, and constructs them into reconstructed attack scenarios. Our evaluation in attack cases shows that TPPR achieves the coordinated optimization of the accuracy and completeness of attack scenario reconstruction through a processing flow of redundant subgraph pruning, path reasoning and screening based on contextual semantic information.


%%\appendix
%%\section{Example Appendix Section}
%%\label{app1}
%%Appendix text.

%%       \citet{<label>} ==> Lamport [21]
%%       \citep{<label>} ==> [21]
%%





% %% For authoryear reference style

% \bibitem[Lamport(1994)]{lamport94}
%   \textit{\LaTeX: a document preparation system},
%   Addison Wesley, Massachusetts,
% \end{thebibliography}
\bibliography{elsarticle-template-num-names} % 确保与.bib文件名一致

\end{document}